\documentclass[aps,prd,reprint,showpacs,amsmath,amssymb,superscriptaddress]{revtex4-2}
\pdfoutput=1
\usepackage{graphicx,hyperref,times,esint}
\hypersetup{
    colorlinks=true,
    linkcolor=blue,
	citecolor=blue,
    filecolor=magenta,      
    urlcolor=cyan
    }

\begin{document}

\title{Acoustic Blackbody Absorption: \\Transcending Causality Limits through Instability-Induced Softness}

\author{Min Yang}
\affiliation{
	Acoustic Metamaterials Group, Data Technology Hub, No. 5 Chun Cheong Street, Hong Kong, China
}
\author{Sichao Qu}
\affiliation{
	Department of Mechanical Engineering, The University of Hong Kong, Pokfulam Road, Hong Kong, China
}
\author{Nicholas Fang}
\affiliation{
	Department of Mechanical Engineering, The University of Hong Kong, Pokfulam Road, Hong Kong, China
}
\author{Shuyu Chen}
\affiliation{
	Acoustic Metamaterials Group, Data Technology Hub, No. 5 Chun Cheong Street, Hong Kong, China
}

\begin{abstract}
By coupling unstable components, we demonstrate a novel approach that reduces static modulus to zero, eliminating causality-imposed absorption limitations in acoustics. Our heuristic model simulations achieve ultra-broadband absorption over 99\% for wavelengths greater than 132 times the absorber thickness. Theoretical analysis further proves this strategy can approach ideal blackbody behavior with infinitesimal thickness. These findings suggest fundamental physical laws no longer prevent true blackbody absorption realization; the only remaining obstacle is the material limitations.
\end{abstract}

\maketitle

The blackbody, an ideal absorber 100\% absorbing wave energy across all wavelengths, has been a fundamental concept in physics since Gustav Kirchhoff's proposal \cite{kirchhoff1860relation}. Kirchhoff and Max Planck's work on blackbody radiation laid the foundation for modern thermodynamics and quantum mechanics, with the blackbody remaining both a theoretical ideal and the ultimate goal in absorption research. This pursuit has driven the development of diverse materials and structures across optics and acoustics. Examples include carbon nanotubes \cite{yang2008experimental, mizuno2009black}, coherent perfect absorbers \cite{chong2010coherent, baranov2017coherent, wang2021coherent, slobodkin2022massively}, and composite metamaterials \cite{landy2008perfect, yu2019broadband, zhou2021ultra, qu2021conceptual, qu2022microwave} for electromagnetic waves, as well as porous materials \cite{allard2009propagation} and resonator-based metamaterials \cite{mei2012dark, yang2017sound, huang2023sound, huang2024acoustic} for sound.

Despite these advancements, scientists have long recognized the practical impossibility of realizing an ideal blackbody. As Planck noted \cite{planck1914theory} that ``all approximately black surfaces which may be realised in practice show appreciable reflection for rays of sufficiently long wave lengths.'' This limitation stems from a universal constraint imposed by the causality principle \cite{rozanov2000ultimate, yang2017optimal, acher2009fundamental, meng2022fundamental}, applicable across all wave types and materials. Recent efforts to overcome these constraints have explored active control schemes \cite{sergeev2023ultrabroadband, wang2024breaking}, time-variance \cite{chen2013broadening, shlivinski2018beyond, li2019beyond, guo2020improving, solis2021functional, li2021temporal, hayran2021spectral, firestein2022absorption, yang2022broadband, hayran2024beyond}, and relaxed boundary conditions \cite{mak2021going, qu2022underwater, firestein2023sum}. However, true blackbody absorption remains elusive. To understand this fundamental limitation and provide potential solutions, we must first examine the causality constraint on absorption in detail.

\begin{figure}
	\includegraphics[scale=0.9]{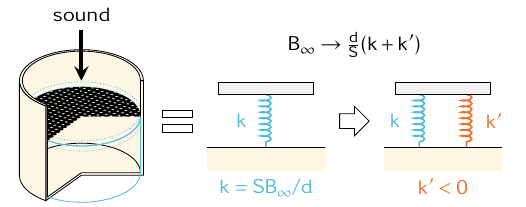}
	\caption{\label{fig:control_modulus}
	Instability raises absorption limit. Cavity air behind MPP acts as a spring (stiffness $k=SB_\infty/d$, $B_\infty$: air's static modulus, $S$: area, $d$: depth). An unstable component (negative spring) reduces effective $B_\infty$, increasing absorption's causal limit, $4\pi^2d\rho c^2/B_\infty$.}
\end{figure}

To do so, consider a classic acoustic micro-perforated plate (MPP) absorber: a rigid plate with micron-sized holes backed by a closed cavity (Fig.~\ref{fig:control_modulus}). When subjected to an external sound pressure $p$ with wavelength $\lambda$, the air in the cavity expands and compresses (blue), driving air molecules through the micropores in a piston-like motion and dissipating sound energy via reciprocating friction. We characterize this system using two parameters: the MPP's acoustic impedance $Z_\text{mpp}=(p-p')/\bar{v}$, where $p'$ is the sound pressure behind the MPP and $\bar{v}$ is the surface-averaged air velocity through the holes; and the cavity's effective bulk modulus $B_\text{eff}=-p'V/\delta V$, with $\delta V$ being the change of the cavity's original volume $V$. These parameters determine the system's absorption ratio:
\begin{equation}
\label{eq:absorption}
	a(\lambda)=1-\left|\frac{B_\text{eff}(\lambda)\lambda+i2\pi dc[\rho c-Z_\text{mpp}(\lambda)]}
	{B_\text{eff}(\lambda)\lambda-i2\pi dc[\rho c+Z_\text{mpp}(\lambda)]}\right|^2,
\end{equation}
where $\rho$ and $c$ are air density and sound speed, and $d$ is the cavity thickness.

To quantify overall absorption, we introduce $\Sigma\equiv-\int_0^\infty\ln[1-a(\lambda)]d\lambda$, representing the total energy loss across the spectrum. For thin cavities and negligible plate thickness, approximating $Z_\text{mpp}(\lambda)$ and $B_\text{eff}(\lambda)$ by their long-wave (static) limit values $Z_\infty$ and $B_\infty$ yields:
\begin{equation}
\label{eq:MPP_sigma}
	\Sigma=\frac{2\pi^2dc}{B_\infty}\left(Z_\infty+\rho c-\left|Z_\infty-\rho c\right|\right).
\end{equation}
A limitation then emerges: as $\Sigma$ increases with $Z_\infty$, a saturated value $4\pi^2d\rho c^2/B_\infty$ exists when $Z_\infty\geq\rho c$. Even though Eq.~\eqref{eq:MPP_sigma} is derived from a specific example, its limit is general. Our previous studies have demonstrated that due to the causal nature of materials being able to absorb sound only from the past, but not from the future, a similar inequality applies to all passive absorbers \cite{yang2017sound,yang2017optimal}.
\begin{equation}
	\label{eq:causality}
	\frac{\Sigma}{4\pi^2d}\leq\frac{\rho c^2}{B_\infty}. 
\end{equation}
This inequality reveals a profound implication: an ideal blackbody cannot exist at a finite thickness because its perfect absorption ($a=1$) for all wavelengths yields an infinite $\Sigma$, which violates the causality constraint.

\begin{figure*}
	\includegraphics[scale=0.9]{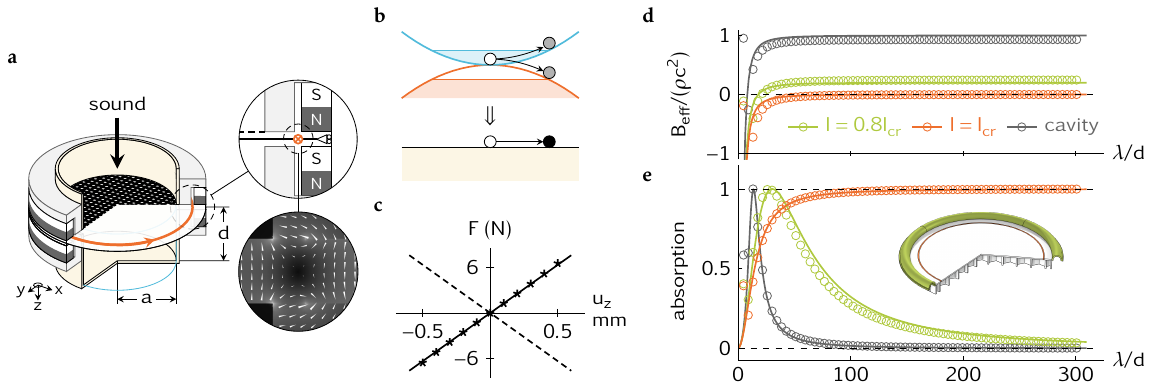}
	\caption{\label{fig:absorption}
	Magnetic quadrupole enhances absorption. (a) Proposed absorber: movable thin plate behind MPP; ring magnets and iron create quadrupole field, inducing vertical instability in internal current (red). (b) Mechanism: unstable energy $-\rho_0\tilde\omega_0^2u_0^2$ (red) cancels original elasticity $(\rho_0\omega_0^2+\rho c^2\Xi_0^2/V)u_0^2$ (blue); $u_0=\int u_l\phi_l^{(0)}dV$, $\Xi_0=\oint\phi_l^{(0)}dS_l$. (c) Simulated Lorentz force at $I_\text{cr}$ (stars) vs. theory (solid); cavity force (dashed). (d,e) Simulated $B_\text{eff}/(\rho c^2)$ and absorption spectra (circles) vs. theory (curves) for various currents; $\lambda/d$: normalized wavelength. Inset: Composite plate geometry for piston-like motion.
	}
\end{figure*}

Another key insight from Eq.~\eqref{eq:causality} is that lowering $B_\infty$ increases the absorption limit. However, reducing the elastic modulus presents significant challenges. For gas in a cavity, $B_\infty$ depends on the thermodynamic process. In typical near-adiabatic conditions, $B_\infty=\gamma p_\text{atm}$, where $\gamma$ is the gas' adiabatic index and $p_\text{atm}$ is atmospheric pressure. Effective heat exchange can decrease $B_\infty$ to $p_\text{atm}$ under isothermal conditions, but this is the lower limit for gas-filled cavities. Adding solid phases inevitably increases the modulus. While opening the back cavity could theoretically achieve zero $B_\infty$, it requires an impractically large space \cite{mak2021going}.

Recent studies on negative and zero stiffness in metastructures \cite{lakes2001extreme, churchill2016dynamically, surjadi2019mechanical, dykstra2023buckling, hussein2024near} offer a promising approach to reduce $B_\infty$, potentially leading to higher absorption \cite{zhao2017membrane, zhang2014thin}. Building on this, we propose coupling the system to an unstable component (Fig.~\ref{fig:control_modulus}). In the following, we will theoretically demonstrate how this approach reduces the static bulk modulus $B_\infty$ and establish requirements for achieving $B_\infty=0$, thereby eliminating the causality constraint on absorption.

To introduce instability, we use an elastic plate sealing the back cavity behind the MPP, with a ring-shaped wire attached to the plate (Fig.~\ref{fig:absorption}a). Placing the wire in a quadrupole magnetic field adds an unstable term, $-\kappa_{il}u_iu_l$, to the plate's potential energy: $E=\int(\lambda_{iklm}\partial_ku_i\partial_mu_l-\kappa_{il}u_iu_l)dV$, where $\lambda_{iklm}$ is the material's elastic modulus tensor, $\partial_k=\partial/\partial x_k$, and $u_l$ is the local displacement. For harmonic oscillations with frequency $\omega=2\pi c/\lambda$, the motion is governed by $\{\omega^2\rho_\text{pl}(x)\delta_{il}+\partial_k[\lambda_{iklm}(x)\partial_m]+\kappa_{il}(x)\}u_l(x)=0$, where $\rho_\text{pl}$ is the plate's density. The relevant Green's function, $g_{lq}(x,x')$, can be expressed using eigenmodes under Neumann boundary condition, $\phi_l^{(n)}(x)$ \cite{yang2014homogenization, yang2014metamaterial}:
\begin{equation}
	\label{eq:green_function}
	g_{lq}(x,x')=\sum_n\frac{\phi_l^{(n)}(x)\phi_q^{(n)}(x')}{\rho_n(\omega_n^2-\tilde\omega_n^2-\omega^2)},
\end{equation}
where the summation is over all the eigenmodes, $\rho_n=\int\phi_i^{(n)}\rho_\text{pl}\delta_{il}\phi_l^{(n)}dV$, $\omega_n^2=-\int\phi_i^{(n)}\partial_k[\lambda_{iklm}\partial_m]\phi_l^{(n)}dV/\rho_n$, and $\tilde\omega_n^2=\int\phi_i^{(n)}\kappa_{il}\phi_l^{(n)}dV/\rho_n$.

With applied pressure $p'$, the displacement is $u_l(x)=\oint g_{lq}(x,x')[p'-p_\text{cavity}(x')]dS'_q$. Given the cavity's reaction pressure $p_\text{cavity}=-\rho c^2\delta V/V$ for long waves under adiabatic condition, solving for $\delta V=-\oint u_ldS_l$ and substituting into $B_\text{eff}$'s definition yields
\begin{equation}
	\label{eq:effective_modulus}
	B_\text{eff}=-p'\frac{V}{\delta V}=\rho c^2+V\left[\oiint g_{lq}(x,x')dS_ldS'_q\right]^{-1}.
\end{equation}
For frequencies near $\sqrt{\omega_0^2-\tilde{\omega}_0^2}$, dominated by one eigenmode, $\phi_l^{(0)}$, this simplifies to
\begin{equation}
	\label{eq:effective_modulus_2}
	B_\text{eff}\simeq \rho c^2+V\rho_0\frac{\omega_0^2-\tilde\omega_0^2-\omega^2}{(\oint\phi_l^{(0)}dS_l)^2}.
\end{equation} 
$B_\text{eff}$ can be reduced to zero at zero-frequency, so that $B_\infty=B_\text{eff}|_{\omega=0}=0$, when:
\begin{equation}
	\label{eq:critical_condition}
	\tilde\omega_0^2=\omega_0^2+\frac{\rho c^2}{V\rho_0}\left(\oint\phi_l^{(0)}dS_l\right)^2,
\end{equation}
resulting in
\begin{equation}
	\label{eq:effective_modulus_3}
	B_\text{eff}=-\omega^2V\frac{\int(\phi_l^{(0)})^2\rho dV}{(\oint\phi_l^{(0)}dS_l)^2}.
\end{equation}

Equation~\eqref{eq:critical_condition} indicates that the unstable potential' 0th eigencomponent cancels out the component of system's elasticity at zero frequency. As shown in Fig.~\ref{fig:absorption}b, the energy increase in the elastic system during the deformation (blue curve) is compensated by the unstable component (red curve), resulting in no apparent energy increase externally. This soft cavity ($B_\text{eff}=0$) allows full absorption when $Z_\text{mpp}=\rho c$, according to Eq.~\eqref{eq:absorption}.

However, Eq.~\eqref{eq:effective_modulus_3} does not equal 0 at non-zero frequencies, and the absorption will drop rapidly as it moves away from the static limit. To approach an ideal blackbody, we must minimize the fractional term in Eq.~\eqref{eq:effective_modulus_3}, necessitating a normalized $\phi_l^{(0)}=1/\sqrt{S\varepsilon}z_l$ representing a piston-like motion for the plate, where $S$ and $\varepsilon$ are the plate's surface area and thickness, and $z_l$ is the unit normal direction. An ideal plate should slide freely in the cavity with sufficient rigidity.

For such piston-like motion, $\omega_0=0$ and $B_\text{eff}=-\omega^2dm/S$, where $m=\int\rho_\text{pl}dV$ is the plate's mass. A closer approximation to an ideal blackbody requires a lighter plate or a shallower cavity. The soft condition in Eq.~\eqref{eq:critical_condition} becomes $\tilde\omega_0^2=\int\kappa_{ii}dV/m=\rho c^2S/(md)$. Since the unstable field exerts a force $F_i=dE/du_i=-u_i\int\kappa_{ii}dV$ on the plate, it is equivalent to a force balance:
\begin{equation}
	\label{eq:critical_condition_2}
	F_i=-u_i\rho c^2S/d=-p_\text{cavity}S=-F_\text{cavity},
\end{equation}
showing that the unstable force precisely offsets the force from the rear cavity.

The key lies in constructing an unstable external field and effectively coupling it with the system. As illustrated in Fig.~\ref{fig:absorption}a, a pair of ring magnets with the same pole orientation are placed outside the cavity, separated by a thin gap. Two L-shaped iron rings direct the magnetic poles to the middle gap, forming a ring-shaped belt of magnetic quadrupoles \cite{hytch2022quadrupoles}. A coil on the extended piston-plate carrying the arrow indicated current (red circle) remains in equilibrium at the middle position, but any slight vertical displacement induces a Lorentz force amplifying the displacement, creating an unstable equilibrium state. The unstable Lorentz force is proportional to the current, $I$, allowing fine-tuning to achieve the soft condition given by Eq.~\eqref{eq:critical_condition_2}. Defining $I_\text{cr}$ as the current satisfying the soft boundary condition and $\eta=I/I_\text{cr}$, then $\tilde\omega_0^2=\eta \rho c^2S/(md)$ and, according to Eq.~\eqref{eq:effective_modulus_2},
\begin{equation}
	\label{eq:effective_modulus_4}
	B_\text{eff}=(1-\eta)\rho c^2-\omega^2md/S.
\end{equation}
Although the plate itself is unstable, \emph{the entire system remains stable} when coupled with the back cavity, provided that $I<I_\text{cr}$. Therefore, we will avoid $I>I_\text{cr}$ causing system unstable.

Full-wave simulations using COMSOL Multiphysics validated our design. The cavity (radius $a=25$ mm, depth $d=65$ mm) has a rigid inner surface. To approximate piston-like motion, a PMMA thin plate is reinforced by ribs and connected to a wrinkled soft PDMS ring sealing the cavity (inset of Fig.~\ref{fig:absorption}e and detailed in Appendix~\ref{sec:numerical_models}). The coil (diameter $0.5$ mm, 10 turns) and the plate have a total mass of $749$ mg. The magnets (remanence $1.2$ T) are separated by a $2$ mm gap. Simulations reveal a well-defined quadrupole magnetic field (Fig.~\ref{fig:absorption}a). A current of $15.5$ A closely achieves the condition described in Eq.~\eqref{eq:critical_condition_2}, with the resulting unstable force shown in Fig.~\ref{fig:absorption}c. Simulated results (stars) deviate slightly from expectations (solid line) for large displacements due to nonlinearity.

Figure~\ref{fig:absorption}d compares the effective modulus with that of the original cavity. As the current increases, $B_\text{eff}$ decreases until reaching near-zero values over a wide band at $I_\text{cr}=15.5$ A. Simulation results agree well with Eq.~\eqref{eq:effective_modulus_4}, with minor differences for short waves due to omitted high-order cavity modes.

The MPP provides an acoustic impedance with low dispersion (detailed in Appendix~\ref{sec:micro-perforated_plate}):
\begin{align}
	Z_\text{mpp}&\simeq\frac{32\nu\rho\tau}{\ell^2\varphi}
	-i\frac{\rho(51\ell+80\tau)}{60\varphi}\omega+\frac{\ell^2\rho\tau}{576\nu\varphi}\omega^2,
	\label{eq:mpp_impedance}
\end{align}
where, $\tau$ is thickness, $\ell$ is pore diameter, $\varphi$ is porosity, and $\nu$ is air's kinematic viscosity. We set $\tau=0.5$ mm, $\ell=0.3$ mm, and $\varphi=0.78\%$ to ensure that the first term equals $\rho c$. Fig.~\ref{fig:absorption}e shows the absorption bandwidth broadening with increasing current, reaching 100\% at $\lambda\to\infty$ when $I=I_\text{cr}$. Notably, absorption increases with wavelength, contrary to traditional materials. When $\lambda>63d$, absorption exceeds 95\%, and when $\lambda>132d$, it surpasses 99\%.

\begin{figure}
	\includegraphics[scale=0.9]{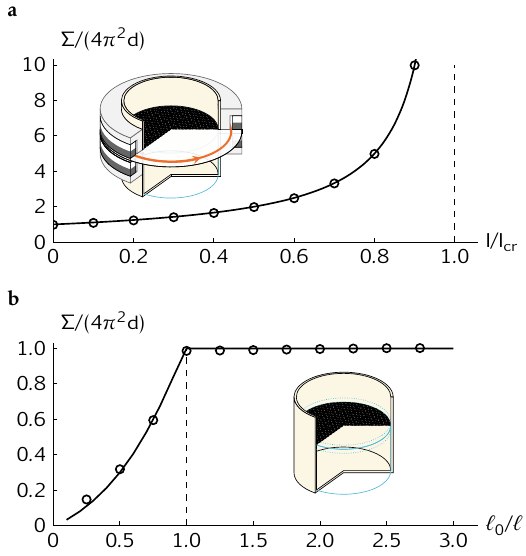}
	\caption{\label{fig:causality}
	Beyond causality constraint. (a) Normalized absorption integral $\Sigma/(4\pi^2d)$ vs. current $I$. $I_\text{cr}$: critical current. (b) $\Sigma/(4\pi^2d)$ vs. MPP pore size $\ell$ for the same cavity and MPP. $\ell_0$: pore diameter in our design. Circles: simulation results; curves: theoretical predictions.}
\end{figure}

Figure~\ref{fig:causality}a examines the causality constraint, showing $\Sigma$ increasing without bound as current rises, ultimately diverging at $I_\text{cr}$—a consequence of $B_\infty=0$. In contrast, traditional cavity-backed MPP absorption improves with reduced perforation size but is limited by a maximum value predicted by Eq.~\eqref{eq:causality}. For pore sizes smaller than a critical $\ell_0$, $\Sigma$ saturates at this value (Fig.~\ref{fig:causality}b, first observed in Ref.~\cite{yang2017sound}). Circles represent simulation data, while curves are derived from Eq.~\eqref{eq:absorption}. The consistency between these results validates the theory's accuracy.

\begin{figure}
	\includegraphics[scale=0.9]{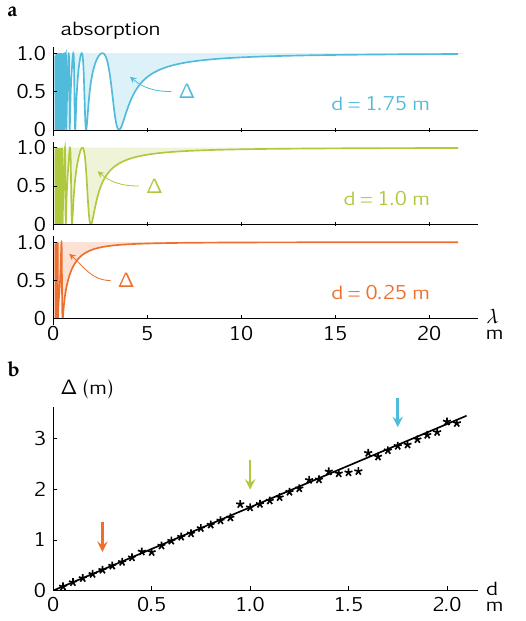}
	\caption{\label{fig:blackbody}
	Towards an acoustic blackbody. (a) Absorption spectra for various sample thicknesses $d$; shaded area $\Delta$: deviation from ideal blackbody. (b) Simulated (stars) and theoretical (line) $\Delta$ vs. $d$. $\Delta$ approaches zero as $d$ decreases.}
\end{figure}

Despite the improvements, absorption at the critical current $I_\text{cr}$ still deteriorates rapidly for short waves. This raises two questions: Can we further enhance absorption to approach that of a blackbody, and are there other limiting principles? To address these, we rewrite Eq.~\eqref{eq:effective_modulus_4} as:	
\begin{equation}
	\label{eq:effective_modulus_5}
	B_\text{eff}=\left[\frac{\omega d}{c}\cot\left(\frac{\omega d}{c}\right)-\eta\right]
	\rho c^2-\omega^2\frac{md}{S}
\end{equation}
to include the cavity's high-order modes, capturing the short wave behaviors. Setting $m=0$, $\eta=1$, and $Z_\text{mpp}=\rho c$ for simplicity, Eq.~\eqref{eq:absorption} becomes:
\begin{equation*}
	a=\left[1+\left(\frac{\lambda}{4\pi d}\right)^2+\frac{1}{4}\cot^2\left(\frac{2\pi d}{\lambda}\right)-\frac{\lambda}{4\pi d}\cot\left(\frac{2\pi d}{\lambda}\right)\right]^{-1}.
\end{equation*}

We quantify the deviation from ideal blackbody absorption using $\Delta$, the shaded areas in Fig.~\ref{fig:blackbody}a. A smaller $\Delta$ indicates closer resemblance to the ideal blackbody. Mathematically:
\begin{equation}
	\label{eq:blackbody_deviation}
	\Delta=\int_0^\infty[1-a(\lambda)]d\lambda=1.64d,
\end{equation}
as shown by the solid line in Fig.~\ref{fig:blackbody}b. As $d$ approaches zero, $a(\lambda)$ converges to the ideal blackbody. However, this improvement comes at a cost: the required current diverges as $I_\text{cr}\sim1/d$. This linear relationship is corroborated by numerical simulations (asterisks in the figure).

In conclusion, we have theoretically and numerically demonstrated that introducing instability can effectively reduce the bulk modulus of acoustic systems, significantly weakening causality-imposed limitations on absorption. By approaching zero static bulk modulus, we realize a soft condition that eliminates the causality constraint. Our thin plate-sealed air cavity model proves that overcoming this limitation substantially broadens the absorption bandwidth. Importantly, our findings reveal that approaching ideal blackbody absorption faces no fundamental physical limitations; the primary obstacles are practical, such as material constraints in tolerating substantial current.

While our study focused on a specific acoustic system, the underlying principles may have broader applications. Given the generality of the causality constraint \cite{qu2022microwave}, the concept of using instability to manipulate material properties could potentially be extended to other fields, such as electromagnetic wave absorption, by achieving high effective magnetic permeability. This approach might offer new perspectives in the design of absorbers and metamaterials across different physical domains.

\appendix

\section{Numerical Models}
\label{sec:numerical_models}

Our numerical models integrate the unstable Lorentz force from a quadrupole magnetic field with a cavity sealed by a composite plate designed for piston-like motion. We employ finite element method (FEM) via COMSOL Multiphysics to solve the relevant differential equations, encompassing both electromagnetic and acoustic components.

The electromagnetic simulation solves static Maxwell equations for permanent magnets and soft iron. We calculate the Lorentz force as ${\bf F}=L\oint{\bf J}\times{\bf B}dS$, where $L$ is coil length, $\bf J$ is current density, and $\bf B$ is the magnetic field vector, integrated over the coils' cross section. As shown in Fig.~\ref{fig:absorption}a, for coils in the quadrupole field's symmetry plane, $\bf F$ is vertical. The key parameter $\int\kappa_{ii}dV=-F_z/u_z$ is determined by calculating $F_z$ for a small vertical displacement $u_z$. This parameter serves as the crucial link between the electromagnetic and acoustic simulations.

In the acoustic model, we directly utilize the $-\int\kappa_{ii}dV$ value obtained from the electromagnetic simulation as the stiffness coefficient of a virtual spring attached on the coil. This approach allows us to incorporate the electromagnetic instability into the acoustic simulation without directly coupling the full electromagnetic equations.

\begin{figure}
    \includegraphics[]{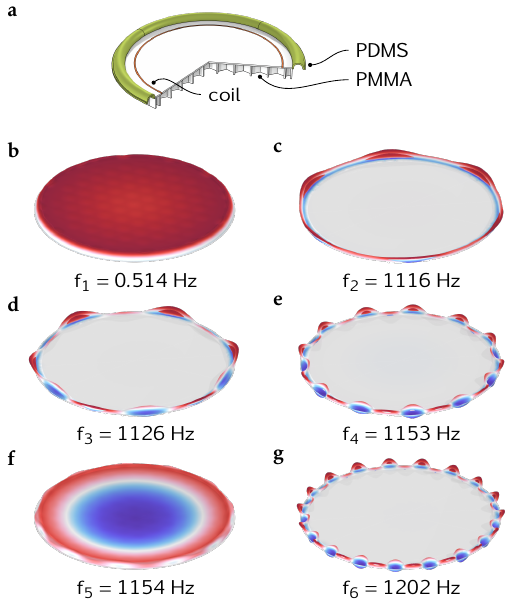}
    \caption{\label{fig:modes}
    Eigenmodes. (a) Schematics of the composite plate. (b-g) First 6 eigenmodes and eigenfrequencies of the system, comprising the composite plate, back cavity, and unstable magnetic force (modeled as a spring foundation with negative spring constant).}
\end{figure}

To achieve uniform piston-like motion despite this localized force, we designed a composite plate structure (Fig.~\ref{fig:modes}a). A PMMA plate is reinforced with a hexagonal lattice frame to enhance rigidity. Additionally, a soft PDMS wrinkled ring on the plate's edge allows free movement along the cavity without significant restoring forces from the fixed edges.

To validate this design, we analyzed the eigenmodes of the complete system (composite plate, back cavity, and virtual spring) with free boundary conditions on the plate's outer surface, as shown in Figure~\ref{fig:modes}.

The results reveal a first eigenmode with a near-zero frequency (0.514 Hz). This extremely low frequency approximates the theoretically expected zero-frequency mode, which arises from the neutral equilibrium state illustrated in Fig.~\ref{fig:absorption}b. This near-zero mode confirms that our system closely achieves the desired cancellation of elasticity by instability.

In this first mode, the composite plate's motion closely approximates piston-like behavior. Higher eigenmodes ($>$1100 Hz) couple weakly with external far-field sounds, further supporting our design's effectiveness. This piston-like approximation over a wide frequency band underpins the accuracy of our theoretical predictions. 

\section{Micro-Perforated Plate (MPP)}
\label{sec:micro-perforated_plate}

The proposed broadband absorption relies on the low dispersion impedance of MPP. An MPP can be modeled as a lattice of short, narrow tubes distributed on a sound-opaque matrix. For a tube with length much shorter than the wavelength, the equation of aerial motion is \cite{rayleigh1929theory-s, crandall1926theory-s}:
\begin{equation}
    -i\omega\rho v(r)-
    \frac{\rho\nu}{r}\frac{\partial}{\partial r}
    \left[r\frac{\partial v(r)}{\partial r}\right]
    =\frac{\Delta p}{\tau},
\end{equation}
where $\Delta p$ is the sound pressure difference across the tube and $r$ is the radial coordinate. The solution for particle velocity $v(r)$, considering viscosity effects at the tube surface ($v=0$ at $r=\ell/2$), is:
\begin{equation}
    v(r)=\frac{i\Delta p}{\tau\rho\omega}\left[
    1-\frac{J_0\left(\sqrt{i\omega/\nu}r\right)}{J_0\left(\sqrt{i\omega/\nu}\ell/2\right)}
    \right].
\end{equation}

For pore diameters and inter-pore distances small compared to the wavelength, the MPP's acoustic properties are described by impedance $Z_\text{MPP}$:
\begin{align}
	Z_\text{MPP}&\equiv\frac{\Delta p}{\bar v}
	=\frac{\Delta p\oint dA}{\varphi\oint vdA}\nonumber\\
	&=-i\omega\frac{\rho\tau}{\varphi}\frac{J_0[\sqrt{i\omega/(4\nu)}\ell]}{J_2[\sqrt{i\omega/(4\nu)}\ell]}-i0.85\omega\frac{\rho\ell}{\varphi}\nonumber\\
	&\simeq32\frac{\nu\rho\tau}{\ell^2\varphi}
	-i\frac{51\ell+80\tau}{60\varphi}\rho\omega+\frac{\ell^2\tau}{576\nu\varphi}\rho\omega^2\cdots.
	\label{eq:S-mpp_impedance}
\end{align}
Here, the term $-i0.85\omega\rho\ell/\varphi$ is Ingard's correction for air motion near tube ends \cite{ingard1953theory-s}. The real 0th-order term indicates constant dissipation over all frequencies. Low dispersion is achieved with small $\ell$, maximizing the contrast between the 0th- and highe-order terms. The consistency between the numerical simulation results based on thermoviscous acoustics shown in Fig.~\ref{fig:causality}b and the predictions derived from Eq.~\eqref{eq:S-mpp_impedance} validates the accuracy of our impedance expression.

\bibliography{References.bib}

\begin{thebibliography}{50}%
\makeatletter
\providecommand \@ifxundefined [1]{%
 \@ifx{#1\undefined}
}%
\providecommand \@ifnum [1]{%
 \ifnum #1\expandafter \@firstoftwo
 \else \expandafter \@secondoftwo
 \fi
}%
\providecommand \@ifx [1]{%
 \ifx #1\expandafter \@firstoftwo
 \else \expandafter \@secondoftwo
 \fi
}%
\providecommand \natexlab [1]{#1}%
\providecommand \enquote  [1]{``#1''}%
\providecommand \bibnamefont  [1]{#1}%
\providecommand \bibfnamefont [1]{#1}%
\providecommand \citenamefont [1]{#1}%
\providecommand \href@noop [0]{\@secondoftwo}%
\providecommand \href [0]{\begingroup \@sanitize@url \@href}%
\providecommand \@href[1]{\@@startlink{#1}\@@href}%
\providecommand \@@href[1]{\endgroup#1\@@endlink}%
\providecommand \@sanitize@url [0]{\catcode `\\12\catcode `\$12\catcode
  `\&12\catcode `\#12\catcode `\^12\catcode `\_12\catcode `\%12\relax}%
\providecommand \@@startlink[1]{}%
\providecommand \@@endlink[0]{}%
\providecommand \url  [0]{\begingroup\@sanitize@url \@url }%
\providecommand \@url [1]{\endgroup\@href {#1}{\urlprefix }}%
\providecommand \urlprefix  [0]{URL }%
\providecommand \Eprint [0]{\href }%
\providecommand \doibase [0]{https://doi.org/}%
\providecommand \selectlanguage [0]{\@gobble}%
\providecommand \bibinfo  [0]{\@secondoftwo}%
\providecommand \bibfield  [0]{\@secondoftwo}%
\providecommand \translation [1]{[#1]}%
\providecommand \BibitemOpen [0]{}%
\providecommand \bibitemStop [0]{}%
\providecommand \bibitemNoStop [0]{.\EOS\space}%
\providecommand \EOS [0]{\spacefactor3000\relax}%
\providecommand \BibitemShut  [1]{\csname bibitem#1\endcsname}%
\let\auto@bib@innerbib\@empty
\bibitem [{\citenamefont {Kirchhoff}(1860)}]{kirchhoff1860relation}%
  \BibitemOpen
  \bibfield  {author} {\bibinfo {author} {\bibfnamefont {G.}~\bibnamefont
  {Kirchhoff}},\ }\bibfield  {title} {\bibinfo {title} {I. on the relation
  between the radiating and absorbing powers of different bodies for light and
  heat},\ }\href@noop {} {\bibfield  {journal} {\bibinfo  {journal} {The
  London, Edinburgh, and Dublin Philosophical Magazine and Journal of Science}\
  }\textbf {\bibinfo {volume} {20}},\ \bibinfo {pages} {1} (\bibinfo {year}
  {1860})}\BibitemShut {NoStop}%
\bibitem [{\citenamefont {Yang}\ \emph {et~al.}(2008)\citenamefont {Yang},
  \citenamefont {Ci}, \citenamefont {Bur}, \citenamefont {Lin},\ and\
  \citenamefont {Ajayan}}]{yang2008experimental}%
  \BibitemOpen
  \bibfield  {author} {\bibinfo {author} {\bibfnamefont {Z.-P.}\ \bibnamefont
  {Yang}}, \bibinfo {author} {\bibfnamefont {L.}~\bibnamefont {Ci}}, \bibinfo
  {author} {\bibfnamefont {J.~A.}\ \bibnamefont {Bur}}, \bibinfo {author}
  {\bibfnamefont {S.-Y.}\ \bibnamefont {Lin}},\ and\ \bibinfo {author}
  {\bibfnamefont {P.~M.}\ \bibnamefont {Ajayan}},\ }\bibfield  {title}
  {\bibinfo {title} {Experimental observation of an extremely dark material
  made by a low-density nanotube array},\ }\href@noop {} {\bibfield  {journal}
  {\bibinfo  {journal} {Nano letters}\ }\textbf {\bibinfo {volume} {8}},\
  \bibinfo {pages} {446} (\bibinfo {year} {2008})}\BibitemShut {NoStop}%
\bibitem [{\citenamefont {Mizuno}\ \emph {et~al.}(2009)\citenamefont {Mizuno},
  \citenamefont {Ishii}, \citenamefont {Kishida}, \citenamefont {Hayamizu},
  \citenamefont {Yasuda}, \citenamefont {Futaba}, \citenamefont {Yumura},\ and\
  \citenamefont {Hata}}]{mizuno2009black}%
  \BibitemOpen
  \bibfield  {author} {\bibinfo {author} {\bibfnamefont {K.}~\bibnamefont
  {Mizuno}}, \bibinfo {author} {\bibfnamefont {J.}~\bibnamefont {Ishii}},
  \bibinfo {author} {\bibfnamefont {H.}~\bibnamefont {Kishida}}, \bibinfo
  {author} {\bibfnamefont {Y.}~\bibnamefont {Hayamizu}}, \bibinfo {author}
  {\bibfnamefont {S.}~\bibnamefont {Yasuda}}, \bibinfo {author} {\bibfnamefont
  {D.~N.}\ \bibnamefont {Futaba}}, \bibinfo {author} {\bibfnamefont
  {M.}~\bibnamefont {Yumura}},\ and\ \bibinfo {author} {\bibfnamefont
  {K.}~\bibnamefont {Hata}},\ }\bibfield  {title} {\bibinfo {title} {A black
  body absorber from vertically aligned single-walled carbon nanotubes},\
  }\href@noop {} {\bibfield  {journal} {\bibinfo  {journal} {Proceedings of the
  National Academy of Sciences}\ }\textbf {\bibinfo {volume} {106}},\ \bibinfo
  {pages} {6044} (\bibinfo {year} {2009})}\BibitemShut {NoStop}%
\bibitem [{\citenamefont {Chong}\ \emph {et~al.}(2010)\citenamefont {Chong},
  \citenamefont {Ge}, \citenamefont {Cao},\ and\ \citenamefont
  {Stone}}]{chong2010coherent}%
  \BibitemOpen
  \bibfield  {author} {\bibinfo {author} {\bibfnamefont {Y.}~\bibnamefont
  {Chong}}, \bibinfo {author} {\bibfnamefont {L.}~\bibnamefont {Ge}}, \bibinfo
  {author} {\bibfnamefont {H.}~\bibnamefont {Cao}},\ and\ \bibinfo {author}
  {\bibfnamefont {A.~D.}\ \bibnamefont {Stone}},\ }\bibfield  {title} {\bibinfo
  {title} {Coherent perfect absorbers: time-reversed lasers},\ }\href@noop {}
  {\bibfield  {journal} {\bibinfo  {journal} {Physical review letters}\
  }\textbf {\bibinfo {volume} {105}},\ \bibinfo {pages} {053901} (\bibinfo
  {year} {2010})}\BibitemShut {NoStop}%
\bibitem [{\citenamefont {Baranov}\ \emph {et~al.}(2017)\citenamefont
  {Baranov}, \citenamefont {Krasnok}, \citenamefont {Shegai}, \citenamefont
  {Al{\`u}},\ and\ \citenamefont {Chong}}]{baranov2017coherent}%
  \BibitemOpen
  \bibfield  {author} {\bibinfo {author} {\bibfnamefont {D.~G.}\ \bibnamefont
  {Baranov}}, \bibinfo {author} {\bibfnamefont {A.}~\bibnamefont {Krasnok}},
  \bibinfo {author} {\bibfnamefont {T.}~\bibnamefont {Shegai}}, \bibinfo
  {author} {\bibfnamefont {A.}~\bibnamefont {Al{\`u}}},\ and\ \bibinfo {author}
  {\bibfnamefont {Y.}~\bibnamefont {Chong}},\ }\bibfield  {title} {\bibinfo
  {title} {Coherent perfect absorbers: linear control of light with light},\
  }\href@noop {} {\bibfield  {journal} {\bibinfo  {journal} {Nature Reviews
  Materials}\ }\textbf {\bibinfo {volume} {2}},\ \bibinfo {pages} {1} (\bibinfo
  {year} {2017})}\BibitemShut {NoStop}%
\bibitem [{\citenamefont {Wang}\ \emph {et~al.}(2021)\citenamefont {Wang},
  \citenamefont {Sweeney}, \citenamefont {Stone},\ and\ \citenamefont
  {Yang}}]{wang2021coherent}%
  \BibitemOpen
  \bibfield  {author} {\bibinfo {author} {\bibfnamefont {C.}~\bibnamefont
  {Wang}}, \bibinfo {author} {\bibfnamefont {W.~R.}\ \bibnamefont {Sweeney}},
  \bibinfo {author} {\bibfnamefont {A.~D.}\ \bibnamefont {Stone}},\ and\
  \bibinfo {author} {\bibfnamefont {L.}~\bibnamefont {Yang}},\ }\bibfield
  {title} {\bibinfo {title} {Coherent perfect absorption at an exceptional
  point},\ }\href@noop {} {\bibfield  {journal} {\bibinfo  {journal} {Science}\
  }\textbf {\bibinfo {volume} {373}},\ \bibinfo {pages} {1261} (\bibinfo {year}
  {2021})}\BibitemShut {NoStop}%
\bibitem [{\citenamefont {Slobodkin}\ \emph {et~al.}(2022)\citenamefont
  {Slobodkin}, \citenamefont {Weinberg}, \citenamefont {H{\"o}rner},
  \citenamefont {Pichler}, \citenamefont {Rotter},\ and\ \citenamefont
  {Katz}}]{slobodkin2022massively}%
  \BibitemOpen
  \bibfield  {author} {\bibinfo {author} {\bibfnamefont {Y.}~\bibnamefont
  {Slobodkin}}, \bibinfo {author} {\bibfnamefont {G.}~\bibnamefont {Weinberg}},
  \bibinfo {author} {\bibfnamefont {H.}~\bibnamefont {H{\"o}rner}}, \bibinfo
  {author} {\bibfnamefont {K.}~\bibnamefont {Pichler}}, \bibinfo {author}
  {\bibfnamefont {S.}~\bibnamefont {Rotter}},\ and\ \bibinfo {author}
  {\bibfnamefont {O.}~\bibnamefont {Katz}},\ }\bibfield  {title} {\bibinfo
  {title} {Massively degenerate coherent perfect absorber for arbitrary
  wavefronts},\ }\href@noop {} {\bibfield  {journal} {\bibinfo  {journal}
  {Science}\ }\textbf {\bibinfo {volume} {377}},\ \bibinfo {pages} {995}
  (\bibinfo {year} {2022})}\BibitemShut {NoStop}%
\bibitem [{\citenamefont {Landy}\ \emph {et~al.}(2008)\citenamefont {Landy},
  \citenamefont {Sajuyigbe}, \citenamefont {Mock}, \citenamefont {Smith},\ and\
  \citenamefont {Padilla}}]{landy2008perfect}%
  \BibitemOpen
  \bibfield  {author} {\bibinfo {author} {\bibfnamefont {N.~I.}\ \bibnamefont
  {Landy}}, \bibinfo {author} {\bibfnamefont {S.}~\bibnamefont {Sajuyigbe}},
  \bibinfo {author} {\bibfnamefont {J.~J.}\ \bibnamefont {Mock}}, \bibinfo
  {author} {\bibfnamefont {D.~R.}\ \bibnamefont {Smith}},\ and\ \bibinfo
  {author} {\bibfnamefont {W.~J.}\ \bibnamefont {Padilla}},\ }\bibfield
  {title} {\bibinfo {title} {Perfect metamaterial absorber},\ }\href@noop {}
  {\bibfield  {journal} {\bibinfo  {journal} {Physical review letters}\
  }\textbf {\bibinfo {volume} {100}},\ \bibinfo {pages} {207402} (\bibinfo
  {year} {2008})}\BibitemShut {NoStop}%
\bibitem [{\citenamefont {Yu}\ \emph {et~al.}(2019)\citenamefont {Yu},
  \citenamefont {Besteiro}, \citenamefont {Huang}, \citenamefont {Wu},
  \citenamefont {Fu}, \citenamefont {Tan}, \citenamefont {Jagadish},
  \citenamefont {Wiederrecht}, \citenamefont {Govorov},\ and\ \citenamefont
  {Wang}}]{yu2019broadband}%
  \BibitemOpen
  \bibfield  {author} {\bibinfo {author} {\bibfnamefont {P.}~\bibnamefont
  {Yu}}, \bibinfo {author} {\bibfnamefont {L.~V.}\ \bibnamefont {Besteiro}},
  \bibinfo {author} {\bibfnamefont {Y.}~\bibnamefont {Huang}}, \bibinfo
  {author} {\bibfnamefont {J.}~\bibnamefont {Wu}}, \bibinfo {author}
  {\bibfnamefont {L.}~\bibnamefont {Fu}}, \bibinfo {author} {\bibfnamefont
  {H.~H.}\ \bibnamefont {Tan}}, \bibinfo {author} {\bibfnamefont
  {C.}~\bibnamefont {Jagadish}}, \bibinfo {author} {\bibfnamefont {G.~P.}\
  \bibnamefont {Wiederrecht}}, \bibinfo {author} {\bibfnamefont {A.~O.}\
  \bibnamefont {Govorov}},\ and\ \bibinfo {author} {\bibfnamefont
  {Z.}~\bibnamefont {Wang}},\ }\bibfield  {title} {\bibinfo {title} {Broadband
  metamaterial absorbers},\ }\href@noop {} {\bibfield  {journal} {\bibinfo
  {journal} {Advanced Optical Materials}\ }\textbf {\bibinfo {volume} {7}},\
  \bibinfo {pages} {1800995} (\bibinfo {year} {2019})}\BibitemShut {NoStop}%
\bibitem [{\citenamefont {Zhou}\ \emph {et~al.}(2021)\citenamefont {Zhou},
  \citenamefont {Qin}, \citenamefont {Liang}, \citenamefont {Meng},
  \citenamefont {Xu}, \citenamefont {Smith},\ and\ \citenamefont
  {Liu}}]{zhou2021ultra}%
  \BibitemOpen
  \bibfield  {author} {\bibinfo {author} {\bibfnamefont {Y.}~\bibnamefont
  {Zhou}}, \bibinfo {author} {\bibfnamefont {Z.}~\bibnamefont {Qin}}, \bibinfo
  {author} {\bibfnamefont {Z.}~\bibnamefont {Liang}}, \bibinfo {author}
  {\bibfnamefont {D.}~\bibnamefont {Meng}}, \bibinfo {author} {\bibfnamefont
  {H.}~\bibnamefont {Xu}}, \bibinfo {author} {\bibfnamefont {D.~R.}\
  \bibnamefont {Smith}},\ and\ \bibinfo {author} {\bibfnamefont
  {Y.}~\bibnamefont {Liu}},\ }\bibfield  {title} {\bibinfo {title}
  {Ultra-broadband metamaterial absorbers from long to very long infrared
  regime},\ }\href@noop {} {\bibfield  {journal} {\bibinfo  {journal} {Light:
  Science \& Applications}\ }\textbf {\bibinfo {volume} {10}},\ \bibinfo
  {pages} {138} (\bibinfo {year} {2021})}\BibitemShut {NoStop}%
\bibitem [{\citenamefont {Qu}\ \emph {et~al.}(2021)\citenamefont {Qu},
  \citenamefont {Hou},\ and\ \citenamefont {Sheng}}]{qu2021conceptual}%
  \BibitemOpen
  \bibfield  {author} {\bibinfo {author} {\bibfnamefont {S.}~\bibnamefont
  {Qu}}, \bibinfo {author} {\bibfnamefont {Y.}~\bibnamefont {Hou}},\ and\
  \bibinfo {author} {\bibfnamefont {P.}~\bibnamefont {Sheng}},\ }\bibfield
  {title} {\bibinfo {title} {Conceptual-based design of an ultrabroadband
  microwave metamaterial absorber},\ }\href@noop {} {\bibfield  {journal}
  {\bibinfo  {journal} {Proceedings of the National Academy of Sciences}\
  }\textbf {\bibinfo {volume} {118}},\ \bibinfo {pages} {e2110490118} (\bibinfo
  {year} {2021})}\BibitemShut {NoStop}%
\bibitem [{\citenamefont {Qu}\ and\ \citenamefont
  {Sheng}(2022)}]{qu2022microwave}%
  \BibitemOpen
  \bibfield  {author} {\bibinfo {author} {\bibfnamefont {S.}~\bibnamefont
  {Qu}}\ and\ \bibinfo {author} {\bibfnamefont {P.}~\bibnamefont {Sheng}},\
  }\bibfield  {title} {\bibinfo {title} {Microwave and acoustic absorption
  metamaterials},\ }\href@noop {} {\bibfield  {journal} {\bibinfo  {journal}
  {Physical Review Applied}\ }\textbf {\bibinfo {volume} {17}},\ \bibinfo
  {pages} {047001} (\bibinfo {year} {2022})}\BibitemShut {NoStop}%
\bibitem [{\citenamefont {Allard}\ and\ \citenamefont
  {Atalla}(2009)}]{allard2009propagation}%
  \BibitemOpen
  \bibfield  {author} {\bibinfo {author} {\bibfnamefont {J.}~\bibnamefont
  {Allard}}\ and\ \bibinfo {author} {\bibfnamefont {N.}~\bibnamefont
  {Atalla}},\ }\href@noop {} {\emph {\bibinfo {title} {Propagation of sound in
  porous media: modelling sound absorbing materials}}}\ (\bibinfo  {publisher}
  {John Wiley \& Sons},\ \bibinfo {year} {2009})\BibitemShut {NoStop}%
\bibitem [{\citenamefont {Mei}\ \emph {et~al.}(2012)\citenamefont {Mei},
  \citenamefont {Ma}, \citenamefont {Yang}, \citenamefont {Yang}, \citenamefont
  {Wen},\ and\ \citenamefont {Sheng}}]{mei2012dark}%
  \BibitemOpen
  \bibfield  {author} {\bibinfo {author} {\bibfnamefont {J.}~\bibnamefont
  {Mei}}, \bibinfo {author} {\bibfnamefont {G.}~\bibnamefont {Ma}}, \bibinfo
  {author} {\bibfnamefont {M.}~\bibnamefont {Yang}}, \bibinfo {author}
  {\bibfnamefont {Z.}~\bibnamefont {Yang}}, \bibinfo {author} {\bibfnamefont
  {W.}~\bibnamefont {Wen}},\ and\ \bibinfo {author} {\bibfnamefont
  {P.}~\bibnamefont {Sheng}},\ }\bibfield  {title} {\bibinfo {title} {Dark
  acoustic metamaterials as super absorbers for low-frequency sound},\
  }\href@noop {} {\bibfield  {journal} {\bibinfo  {journal} {Nature
  communications}\ }\textbf {\bibinfo {volume} {3}},\ \bibinfo {pages} {756}
  (\bibinfo {year} {2012})}\BibitemShut {NoStop}%
\bibitem [{\citenamefont {Yang}\ and\ \citenamefont
  {Sheng}(2017)}]{yang2017sound}%
  \BibitemOpen
  \bibfield  {author} {\bibinfo {author} {\bibfnamefont {M.}~\bibnamefont
  {Yang}}\ and\ \bibinfo {author} {\bibfnamefont {P.}~\bibnamefont {Sheng}},\
  }\bibfield  {title} {\bibinfo {title} {Sound absorption structures: From
  porous media to acoustic metamaterials},\ }\href@noop {} {\bibfield
  {journal} {\bibinfo  {journal} {Annual Review of Materials Research}\
  }\textbf {\bibinfo {volume} {47}},\ \bibinfo {pages} {83} (\bibinfo {year}
  {2017})}\BibitemShut {NoStop}%
\bibitem [{\citenamefont {Huang}\ \emph {et~al.}(2023)\citenamefont {Huang},
  \citenamefont {Li}, \citenamefont {Zhu},\ and\ \citenamefont
  {Tsai}}]{huang2023sound}%
  \BibitemOpen
  \bibfield  {author} {\bibinfo {author} {\bibfnamefont {S.}~\bibnamefont
  {Huang}}, \bibinfo {author} {\bibfnamefont {Y.}~\bibnamefont {Li}}, \bibinfo
  {author} {\bibfnamefont {J.}~\bibnamefont {Zhu}},\ and\ \bibinfo {author}
  {\bibfnamefont {D.~P.}\ \bibnamefont {Tsai}},\ }\bibfield  {title} {\bibinfo
  {title} {Sound-absorbing materials},\ }\href@noop {} {\bibfield  {journal}
  {\bibinfo  {journal} {Physical Review Applied}\ }\textbf {\bibinfo {volume}
  {20}},\ \bibinfo {pages} {010501} (\bibinfo {year} {2023})}\BibitemShut
  {NoStop}%
\bibitem [{\citenamefont {Huang}\ \emph {et~al.}(2024)\citenamefont {Huang},
  \citenamefont {Huang}, \citenamefont {Shen}, \citenamefont {Yves},
  \citenamefont {Pilipchuk}, \citenamefont {Ni}, \citenamefont {Kim},
  \citenamefont {Chiang}, \citenamefont {Powell}, \citenamefont {Zhu} \emph
  {et~al.}}]{huang2024acoustic}%
  \BibitemOpen
  \bibfield  {author} {\bibinfo {author} {\bibfnamefont {L.}~\bibnamefont
  {Huang}}, \bibinfo {author} {\bibfnamefont {S.}~\bibnamefont {Huang}},
  \bibinfo {author} {\bibfnamefont {C.}~\bibnamefont {Shen}}, \bibinfo {author}
  {\bibfnamefont {S.}~\bibnamefont {Yves}}, \bibinfo {author} {\bibfnamefont
  {A.~S.}\ \bibnamefont {Pilipchuk}}, \bibinfo {author} {\bibfnamefont
  {X.}~\bibnamefont {Ni}}, \bibinfo {author} {\bibfnamefont {S.}~\bibnamefont
  {Kim}}, \bibinfo {author} {\bibfnamefont {Y.~K.}\ \bibnamefont {Chiang}},
  \bibinfo {author} {\bibfnamefont {D.~A.}\ \bibnamefont {Powell}}, \bibinfo
  {author} {\bibfnamefont {J.}~\bibnamefont {Zhu}}, \emph {et~al.},\ }\bibfield
   {title} {\bibinfo {title} {Acoustic resonances in non-hermitian open
  systems},\ }\href@noop {} {\bibfield  {journal} {\bibinfo  {journal} {Nature
  Reviews Physics}\ }\textbf {\bibinfo {volume} {6}},\ \bibinfo {pages} {11}
  (\bibinfo {year} {2024})}\BibitemShut {NoStop}%
\bibitem [{\citenamefont {Planck}(1914)}]{planck1914theory}%
  \BibitemOpen
  \bibfield  {author} {\bibinfo {author} {\bibfnamefont {M.}~\bibnamefont
  {Planck}},\ }\href@noop {} {\emph {\bibinfo {title} {The theory of heat
  radiation}}}\ (\bibinfo  {publisher} {Philadelphia, P. Blakiston's Son \&
  Co},\ \bibinfo {year} {1914})\BibitemShut {NoStop}%
\bibitem [{\citenamefont {Rozanov}(2000)}]{rozanov2000ultimate}%
  \BibitemOpen
  \bibfield  {author} {\bibinfo {author} {\bibfnamefont {K.~N.}\ \bibnamefont
  {Rozanov}},\ }\bibfield  {title} {\bibinfo {title} {Ultimate thickness to
  bandwidth ratio of radar absorbers},\ }\href@noop {} {\bibfield  {journal}
  {\bibinfo  {journal} {IEEE Transactions on Antennas and Propagation}\
  }\textbf {\bibinfo {volume} {48}},\ \bibinfo {pages} {1230} (\bibinfo {year}
  {2000})}\BibitemShut {NoStop}%
\bibitem [{\citenamefont {Yang}\ \emph {et~al.}(2017)\citenamefont {Yang},
  \citenamefont {Chen}, \citenamefont {Fu},\ and\ \citenamefont
  {Sheng}}]{yang2017optimal}%
  \BibitemOpen
  \bibfield  {author} {\bibinfo {author} {\bibfnamefont {M.}~\bibnamefont
  {Yang}}, \bibinfo {author} {\bibfnamefont {S.}~\bibnamefont {Chen}}, \bibinfo
  {author} {\bibfnamefont {C.}~\bibnamefont {Fu}},\ and\ \bibinfo {author}
  {\bibfnamefont {P.}~\bibnamefont {Sheng}},\ }\bibfield  {title} {\bibinfo
  {title} {Optimal sound-absorbing structures},\ }\href@noop {} {\bibfield
  {journal} {\bibinfo  {journal} {Materials Horizons}\ }\textbf {\bibinfo
  {volume} {4}},\ \bibinfo {pages} {673} (\bibinfo {year} {2017})}\BibitemShut
  {NoStop}%
\bibitem [{\citenamefont {Acher}\ \emph {et~al.}(2009)\citenamefont {Acher},
  \citenamefont {Bernard}, \citenamefont {Mar{\'e}chal}, \citenamefont
  {Bardaine},\ and\ \citenamefont {Levassort}}]{acher2009fundamental}%
  \BibitemOpen
  \bibfield  {author} {\bibinfo {author} {\bibfnamefont {O.}~\bibnamefont
  {Acher}}, \bibinfo {author} {\bibfnamefont {J.}~\bibnamefont {Bernard}},
  \bibinfo {author} {\bibfnamefont {P.}~\bibnamefont {Mar{\'e}chal}}, \bibinfo
  {author} {\bibfnamefont {A.}~\bibnamefont {Bardaine}},\ and\ \bibinfo
  {author} {\bibfnamefont {F.}~\bibnamefont {Levassort}},\ }\bibfield  {title}
  {\bibinfo {title} {Fundamental constraints on the performance of broadband
  ultrasonic matching structures and absorbers},\ }\href@noop {} {\bibfield
  {journal} {\bibinfo  {journal} {The Journal of the Acoustical Society of
  America}\ }\textbf {\bibinfo {volume} {125}},\ \bibinfo {pages} {1995}
  (\bibinfo {year} {2009})}\BibitemShut {NoStop}%
\bibitem [{\citenamefont {Meng}\ \emph {et~al.}(2022)\citenamefont {Meng},
  \citenamefont {Romero-Garc{\'\i}a}, \citenamefont {Gabard}, \citenamefont
  {Groby}, \citenamefont {Bricault}, \citenamefont {Goud{\'e}},\ and\
  \citenamefont {Sheng}}]{meng2022fundamental}%
  \BibitemOpen
  \bibfield  {author} {\bibinfo {author} {\bibfnamefont {Y.}~\bibnamefont
  {Meng}}, \bibinfo {author} {\bibfnamefont {V.}~\bibnamefont
  {Romero-Garc{\'\i}a}}, \bibinfo {author} {\bibfnamefont {G.}~\bibnamefont
  {Gabard}}, \bibinfo {author} {\bibfnamefont {J.-P.}\ \bibnamefont {Groby}},
  \bibinfo {author} {\bibfnamefont {C.}~\bibnamefont {Bricault}}, \bibinfo
  {author} {\bibfnamefont {S.}~\bibnamefont {Goud{\'e}}},\ and\ \bibinfo
  {author} {\bibfnamefont {P.}~\bibnamefont {Sheng}},\ }\bibfield  {title}
  {\bibinfo {title} {Fundamental constraints on broadband passive acoustic
  treatments in unidimensional scattering problems},\ }\href@noop {} {\bibfield
   {journal} {\bibinfo  {journal} {Proceedings of the Royal Society A}\
  }\textbf {\bibinfo {volume} {478}},\ \bibinfo {pages} {20220287} (\bibinfo
  {year} {2022})}\BibitemShut {NoStop}%
\bibitem [{\citenamefont {Sergeev}\ \emph {et~al.}(2023)\citenamefont
  {Sergeev}, \citenamefont {Fleury},\ and\ \citenamefont
  {Lissek}}]{sergeev2023ultrabroadband}%
  \BibitemOpen
  \bibfield  {author} {\bibinfo {author} {\bibfnamefont {S.}~\bibnamefont
  {Sergeev}}, \bibinfo {author} {\bibfnamefont {R.}~\bibnamefont {Fleury}},\
  and\ \bibinfo {author} {\bibfnamefont {H.}~\bibnamefont {Lissek}},\
  }\bibfield  {title} {\bibinfo {title} {Ultrabroadband sound control with
  deep-subwavelength plasmacoustic metalayers},\ }\href@noop {} {\bibfield
  {journal} {\bibinfo  {journal} {Nature Communications}\ }\textbf {\bibinfo
  {volume} {14}},\ \bibinfo {pages} {2874} (\bibinfo {year}
  {2023})}\BibitemShut {NoStop}%
\bibitem [{\citenamefont {Wang}\ \emph {et~al.}(2024)\citenamefont {Wang},
  \citenamefont {Zhao}, \citenamefont {Shen}, \citenamefont {Shi},
  \citenamefont {Zou}, \citenamefont {Lu},\ and\ \citenamefont
  {Al{\`u}}}]{wang2024breaking}%
  \BibitemOpen
  \bibfield  {author} {\bibinfo {author} {\bibfnamefont {K.}~\bibnamefont
  {Wang}}, \bibinfo {author} {\bibfnamefont {S.}~\bibnamefont {Zhao}}, \bibinfo
  {author} {\bibfnamefont {C.}~\bibnamefont {Shen}}, \bibinfo {author}
  {\bibfnamefont {L.}~\bibnamefont {Shi}}, \bibinfo {author} {\bibfnamefont
  {H.}~\bibnamefont {Zou}}, \bibinfo {author} {\bibfnamefont {J.}~\bibnamefont
  {Lu}},\ and\ \bibinfo {author} {\bibfnamefont {A.}~\bibnamefont {Al{\`u}}},\
  }\bibfield  {title} {\bibinfo {title} {Breaking the causality limit for
  broadband acoustic absorption using a noncausal active absorber},\
  }\href@noop {} {\bibfield  {journal} {\bibinfo  {journal} {Device}\ }
  (\bibinfo {year} {2024})}\BibitemShut {NoStop}%
\bibitem [{\citenamefont {Chen}\ \emph {et~al.}(2013)\citenamefont {Chen},
  \citenamefont {Argyropoulos},\ and\ \citenamefont
  {Al{\`u}}}]{chen2013broadening}%
  \BibitemOpen
  \bibfield  {author} {\bibinfo {author} {\bibfnamefont {P.-Y.}\ \bibnamefont
  {Chen}}, \bibinfo {author} {\bibfnamefont {C.}~\bibnamefont {Argyropoulos}},\
  and\ \bibinfo {author} {\bibfnamefont {A.}~\bibnamefont {Al{\`u}}},\
  }\bibfield  {title} {\bibinfo {title} {Broadening the cloaking bandwidth with
  non-foster metasurfaces},\ }\href@noop {} {\bibfield  {journal} {\bibinfo
  {journal} {Physical review letters}\ }\textbf {\bibinfo {volume} {111}},\
  \bibinfo {pages} {233001} (\bibinfo {year} {2013})}\BibitemShut {NoStop}%
\bibitem [{\citenamefont {Shlivinski}\ and\ \citenamefont
  {Hadad}(2018)}]{shlivinski2018beyond}%
  \BibitemOpen
  \bibfield  {author} {\bibinfo {author} {\bibfnamefont {A.}~\bibnamefont
  {Shlivinski}}\ and\ \bibinfo {author} {\bibfnamefont {Y.}~\bibnamefont
  {Hadad}},\ }\bibfield  {title} {\bibinfo {title} {Beyond the bode-fano bound:
  Wideband impedance matching for short pulses using temporal switching of
  transmission-line parameters},\ }\href@noop {} {\bibfield  {journal}
  {\bibinfo  {journal} {Physical review letters}\ }\textbf {\bibinfo {volume}
  {121}},\ \bibinfo {pages} {204301} (\bibinfo {year} {2018})}\BibitemShut
  {NoStop}%
\bibitem [{\citenamefont {Li}\ \emph {et~al.}(2019)\citenamefont {Li},
  \citenamefont {Mekawy},\ and\ \citenamefont {Al{\`u}}}]{li2019beyond}%
  \BibitemOpen
  \bibfield  {author} {\bibinfo {author} {\bibfnamefont {H.}~\bibnamefont
  {Li}}, \bibinfo {author} {\bibfnamefont {A.}~\bibnamefont {Mekawy}},\ and\
  \bibinfo {author} {\bibfnamefont {A.}~\bibnamefont {Al{\`u}}},\ }\bibfield
  {title} {\bibinfo {title} {Beyond chu’s limit with floquet impedance
  matching},\ }\href@noop {} {\bibfield  {journal} {\bibinfo  {journal}
  {Physical review letters}\ }\textbf {\bibinfo {volume} {123}},\ \bibinfo
  {pages} {164102} (\bibinfo {year} {2019})}\BibitemShut {NoStop}%
\bibitem [{\citenamefont {Guo}\ \emph {et~al.}(2020)\citenamefont {Guo},
  \citenamefont {Lissek},\ and\ \citenamefont {Fleury}}]{guo2020improving}%
  \BibitemOpen
  \bibfield  {author} {\bibinfo {author} {\bibfnamefont {X.}~\bibnamefont
  {Guo}}, \bibinfo {author} {\bibfnamefont {H.}~\bibnamefont {Lissek}},\ and\
  \bibinfo {author} {\bibfnamefont {R.}~\bibnamefont {Fleury}},\ }\bibfield
  {title} {\bibinfo {title} {Improving sound absorption through nonlinear
  active electroacoustic resonators},\ }\href@noop {} {\bibfield  {journal}
  {\bibinfo  {journal} {Physical Review Applied}\ }\textbf {\bibinfo {volume}
  {13}},\ \bibinfo {pages} {014018} (\bibinfo {year} {2020})}\BibitemShut
  {NoStop}%
\bibitem [{\citenamefont {Sol{\'\i}s}\ and\ \citenamefont
  {Engheta}(2021)}]{solis2021functional}%
  \BibitemOpen
  \bibfield  {author} {\bibinfo {author} {\bibfnamefont {D.~M.}\ \bibnamefont
  {Sol{\'\i}s}}\ and\ \bibinfo {author} {\bibfnamefont {N.}~\bibnamefont
  {Engheta}},\ }\bibfield  {title} {\bibinfo {title} {Functional analysis of
  the polarization response in linear time-varying media: A generalization of
  the kramers-kronig relations},\ }\href@noop {} {\bibfield  {journal}
  {\bibinfo  {journal} {Physical Review B}\ }\textbf {\bibinfo {volume}
  {103}},\ \bibinfo {pages} {144303} (\bibinfo {year} {2021})}\BibitemShut
  {NoStop}%
\bibitem [{\citenamefont {Li}\ and\ \citenamefont
  {Al{\`u}}(2021)}]{li2021temporal}%
  \BibitemOpen
  \bibfield  {author} {\bibinfo {author} {\bibfnamefont {H.}~\bibnamefont
  {Li}}\ and\ \bibinfo {author} {\bibfnamefont {A.}~\bibnamefont {Al{\`u}}},\
  }\bibfield  {title} {\bibinfo {title} {Temporal switching to extend the
  bandwidth of thin absorbers},\ }\href@noop {} {\bibfield  {journal} {\bibinfo
   {journal} {Optica}\ }\textbf {\bibinfo {volume} {8}},\ \bibinfo {pages} {24}
  (\bibinfo {year} {2021})}\BibitemShut {NoStop}%
\bibitem [{\citenamefont {Hayran}\ \emph {et~al.}(2021)\citenamefont {Hayran},
  \citenamefont {Chen},\ and\ \citenamefont {Monticone}}]{hayran2021spectral}%
  \BibitemOpen
  \bibfield  {author} {\bibinfo {author} {\bibfnamefont {Z.}~\bibnamefont
  {Hayran}}, \bibinfo {author} {\bibfnamefont {A.}~\bibnamefont {Chen}},\ and\
  \bibinfo {author} {\bibfnamefont {F.}~\bibnamefont {Monticone}},\ }\bibfield
  {title} {\bibinfo {title} {Spectral causality and the scattering of waves},\
  }\href@noop {} {\bibfield  {journal} {\bibinfo  {journal} {Optica}\ }\textbf
  {\bibinfo {volume} {8}},\ \bibinfo {pages} {1040} (\bibinfo {year}
  {2021})}\BibitemShut {NoStop}%
\bibitem [{\citenamefont {Firestein}\ \emph {et~al.}(2022)\citenamefont
  {Firestein}, \citenamefont {Shlivinski},\ and\ \citenamefont
  {Hadad}}]{firestein2022absorption}%
  \BibitemOpen
  \bibfield  {author} {\bibinfo {author} {\bibfnamefont {C.}~\bibnamefont
  {Firestein}}, \bibinfo {author} {\bibfnamefont {A.}~\bibnamefont
  {Shlivinski}},\ and\ \bibinfo {author} {\bibfnamefont {Y.}~\bibnamefont
  {Hadad}},\ }\bibfield  {title} {\bibinfo {title} {Absorption and scattering
  by a temporally switched lossy layer: Going beyond the rozanov bound},\
  }\href@noop {} {\bibfield  {journal} {\bibinfo  {journal} {Physical Review
  Applied}\ }\textbf {\bibinfo {volume} {17}},\ \bibinfo {pages} {014017}
  (\bibinfo {year} {2022})}\BibitemShut {NoStop}%
\bibitem [{\citenamefont {Yang}\ \emph {et~al.}(2022)\citenamefont {Yang},
  \citenamefont {Wen},\ and\ \citenamefont {Sievenpiper}}]{yang2022broadband}%
  \BibitemOpen
  \bibfield  {author} {\bibinfo {author} {\bibfnamefont {X.}~\bibnamefont
  {Yang}}, \bibinfo {author} {\bibfnamefont {E.}~\bibnamefont {Wen}},\ and\
  \bibinfo {author} {\bibfnamefont {D.~F.}\ \bibnamefont {Sievenpiper}},\
  }\bibfield  {title} {\bibinfo {title} {Broadband time-modulated absorber
  beyond the bode-fano limit for short pulses by energy trapping},\ }\href@noop
  {} {\bibfield  {journal} {\bibinfo  {journal} {Physical Review Applied}\
  }\textbf {\bibinfo {volume} {17}},\ \bibinfo {pages} {044003} (\bibinfo
  {year} {2022})}\BibitemShut {NoStop}%
\bibitem [{\citenamefont {Hayran}\ and\ \citenamefont
  {Monticone}(2024)}]{hayran2024beyond}%
  \BibitemOpen
  \bibfield  {author} {\bibinfo {author} {\bibfnamefont {Z.}~\bibnamefont
  {Hayran}}\ and\ \bibinfo {author} {\bibfnamefont {F.}~\bibnamefont
  {Monticone}},\ }\bibfield  {title} {\bibinfo {title} {Beyond the rozanov
  bound on electromagnetic absorption via periodic temporal modulations},\
  }\href@noop {} {\bibfield  {journal} {\bibinfo  {journal} {Physical Review
  Applied}\ }\textbf {\bibinfo {volume} {21}},\ \bibinfo {pages} {044007}
  (\bibinfo {year} {2024})}\BibitemShut {NoStop}%
\bibitem [{\citenamefont {Mak}\ \emph {et~al.}(2021)\citenamefont {Mak},
  \citenamefont {Zhang}, \citenamefont {Dong}, \citenamefont {Miura},
  \citenamefont {Iwata},\ and\ \citenamefont {Sheng}}]{mak2021going}%
  \BibitemOpen
  \bibfield  {author} {\bibinfo {author} {\bibfnamefont {H.~Y.}\ \bibnamefont
  {Mak}}, \bibinfo {author} {\bibfnamefont {X.}~\bibnamefont {Zhang}}, \bibinfo
  {author} {\bibfnamefont {Z.}~\bibnamefont {Dong}}, \bibinfo {author}
  {\bibfnamefont {S.}~\bibnamefont {Miura}}, \bibinfo {author} {\bibfnamefont
  {T.}~\bibnamefont {Iwata}},\ and\ \bibinfo {author} {\bibfnamefont
  {P.}~\bibnamefont {Sheng}},\ }\bibfield  {title} {\bibinfo {title} {Going
  beyond the causal limit in acoustic absorption},\ }\href@noop {} {\bibfield
  {journal} {\bibinfo  {journal} {Physical Review Applied}\ }\textbf {\bibinfo
  {volume} {16}},\ \bibinfo {pages} {044062} (\bibinfo {year}
  {2021})}\BibitemShut {NoStop}%
\bibitem [{\citenamefont {Qu}\ \emph {et~al.}(2022)\citenamefont {Qu},
  \citenamefont {Gao}, \citenamefont {Tinel}, \citenamefont {Morvan},
  \citenamefont {Romero-Garc{\'\i}a}, \citenamefont {Groby},\ and\
  \citenamefont {Sheng}}]{qu2022underwater}%
  \BibitemOpen
  \bibfield  {author} {\bibinfo {author} {\bibfnamefont {S.}~\bibnamefont
  {Qu}}, \bibinfo {author} {\bibfnamefont {N.}~\bibnamefont {Gao}}, \bibinfo
  {author} {\bibfnamefont {A.}~\bibnamefont {Tinel}}, \bibinfo {author}
  {\bibfnamefont {B.}~\bibnamefont {Morvan}}, \bibinfo {author} {\bibfnamefont
  {V.}~\bibnamefont {Romero-Garc{\'\i}a}}, \bibinfo {author} {\bibfnamefont
  {J.-P.}\ \bibnamefont {Groby}},\ and\ \bibinfo {author} {\bibfnamefont
  {P.}~\bibnamefont {Sheng}},\ }\bibfield  {title} {\bibinfo {title}
  {Underwater metamaterial absorber with impedance-matched composite},\
  }\href@noop {} {\bibfield  {journal} {\bibinfo  {journal} {Science Advances}\
  }\textbf {\bibinfo {volume} {8}},\ \bibinfo {pages} {eabm4206} (\bibinfo
  {year} {2022})}\BibitemShut {NoStop}%
\bibitem [{\citenamefont {Firestein}\ \emph {et~al.}(2023)\citenamefont
  {Firestein}, \citenamefont {Shlivinski},\ and\ \citenamefont
  {Hadad}}]{firestein2023sum}%
  \BibitemOpen
  \bibfield  {author} {\bibinfo {author} {\bibfnamefont {C.}~\bibnamefont
  {Firestein}}, \bibinfo {author} {\bibfnamefont {A.}~\bibnamefont
  {Shlivinski}},\ and\ \bibinfo {author} {\bibfnamefont {Y.}~\bibnamefont
  {Hadad}},\ }\bibfield  {title} {\bibinfo {title} {Sum rule bounds beyond
  rozanov criterion in linear and time-invariant thin absorbers},\ }\href@noop
  {} {\bibfield  {journal} {\bibinfo  {journal} {Physical Review B}\ }\textbf
  {\bibinfo {volume} {108}},\ \bibinfo {pages} {014308} (\bibinfo {year}
  {2023})}\BibitemShut {NoStop}%
\bibitem [{\citenamefont {Lakes}\ \emph {et~al.}(2001)\citenamefont {Lakes},
  \citenamefont {Lee}, \citenamefont {Bersie},\ and\ \citenamefont
  {Wang}}]{lakes2001extreme}%
  \BibitemOpen
  \bibfield  {author} {\bibinfo {author} {\bibfnamefont {R.~S.}\ \bibnamefont
  {Lakes}}, \bibinfo {author} {\bibfnamefont {T.}~\bibnamefont {Lee}}, \bibinfo
  {author} {\bibfnamefont {A.}~\bibnamefont {Bersie}},\ and\ \bibinfo {author}
  {\bibfnamefont {Y.-C.}\ \bibnamefont {Wang}},\ }\bibfield  {title} {\bibinfo
  {title} {Extreme damping in composite materials with negative-stiffness
  inclusions},\ }\href@noop {} {\bibfield  {journal} {\bibinfo  {journal}
  {Nature}\ }\textbf {\bibinfo {volume} {410}},\ \bibinfo {pages} {565}
  (\bibinfo {year} {2001})}\BibitemShut {NoStop}%
\bibitem [{\citenamefont {Churchill}\ \emph {et~al.}(2016)\citenamefont
  {Churchill}, \citenamefont {Shahan}, \citenamefont {Smith}, \citenamefont
  {Keefe},\ and\ \citenamefont {McKnight}}]{churchill2016dynamically}%
  \BibitemOpen
  \bibfield  {author} {\bibinfo {author} {\bibfnamefont {C.~B.}\ \bibnamefont
  {Churchill}}, \bibinfo {author} {\bibfnamefont {D.~W.}\ \bibnamefont
  {Shahan}}, \bibinfo {author} {\bibfnamefont {S.~P.}\ \bibnamefont {Smith}},
  \bibinfo {author} {\bibfnamefont {A.~C.}\ \bibnamefont {Keefe}},\ and\
  \bibinfo {author} {\bibfnamefont {G.~P.}\ \bibnamefont {McKnight}},\
  }\bibfield  {title} {\bibinfo {title} {Dynamically variable negative
  stiffness structures},\ }\href@noop {} {\bibfield  {journal} {\bibinfo
  {journal} {Science advances}\ }\textbf {\bibinfo {volume} {2}},\ \bibinfo
  {pages} {e1500778} (\bibinfo {year} {2016})}\BibitemShut {NoStop}%
\bibitem [{\citenamefont {Surjadi}\ \emph {et~al.}(2019)\citenamefont
  {Surjadi}, \citenamefont {Gao}, \citenamefont {Du}, \citenamefont {Li},
  \citenamefont {Xiong}, \citenamefont {Fang},\ and\ \citenamefont
  {Lu}}]{surjadi2019mechanical}%
  \BibitemOpen
  \bibfield  {author} {\bibinfo {author} {\bibfnamefont {J.~U.}\ \bibnamefont
  {Surjadi}}, \bibinfo {author} {\bibfnamefont {L.}~\bibnamefont {Gao}},
  \bibinfo {author} {\bibfnamefont {H.}~\bibnamefont {Du}}, \bibinfo {author}
  {\bibfnamefont {X.}~\bibnamefont {Li}}, \bibinfo {author} {\bibfnamefont
  {X.}~\bibnamefont {Xiong}}, \bibinfo {author} {\bibfnamefont {N.~X.}\
  \bibnamefont {Fang}},\ and\ \bibinfo {author} {\bibfnamefont
  {Y.}~\bibnamefont {Lu}},\ }\bibfield  {title} {\bibinfo {title} {Mechanical
  metamaterials and their engineering applications},\ }\href@noop {} {\bibfield
   {journal} {\bibinfo  {journal} {Advanced Engineering Materials}\ }\textbf
  {\bibinfo {volume} {21}},\ \bibinfo {pages} {1800864} (\bibinfo {year}
  {2019})}\BibitemShut {NoStop}%
\bibitem [{\citenamefont {Dykstra}\ \emph {et~al.}(2023)\citenamefont
  {Dykstra}, \citenamefont {Lenting}, \citenamefont {Masurier},\ and\
  \citenamefont {Coulais}}]{dykstra2023buckling}%
  \BibitemOpen
  \bibfield  {author} {\bibinfo {author} {\bibfnamefont {D.~M.}\ \bibnamefont
  {Dykstra}}, \bibinfo {author} {\bibfnamefont {C.}~\bibnamefont {Lenting}},
  \bibinfo {author} {\bibfnamefont {A.}~\bibnamefont {Masurier}},\ and\
  \bibinfo {author} {\bibfnamefont {C.}~\bibnamefont {Coulais}},\ }\bibfield
  {title} {\bibinfo {title} {Buckling metamaterials for extreme vibration
  damping},\ }\href@noop {} {\bibfield  {journal} {\bibinfo  {journal}
  {Advanced Materials}\ }\textbf {\bibinfo {volume} {35}},\ \bibinfo {pages}
  {2301747} (\bibinfo {year} {2023})}\BibitemShut {NoStop}%
\bibitem [{\citenamefont {Hussein}\ \emph {et~al.}(2024)\citenamefont
  {Hussein}, \citenamefont {Wang}, \citenamefont {Amendoeira~Esteves},
  \citenamefont {Kraft},\ and\ \citenamefont {Fariborzi}}]{hussein2024near}%
  \BibitemOpen
  \bibfield  {author} {\bibinfo {author} {\bibfnamefont {H.}~\bibnamefont
  {Hussein}}, \bibinfo {author} {\bibfnamefont {C.}~\bibnamefont {Wang}},
  \bibinfo {author} {\bibfnamefont {R.}~\bibnamefont {Amendoeira~Esteves}},
  \bibinfo {author} {\bibfnamefont {M.}~\bibnamefont {Kraft}},\ and\ \bibinfo
  {author} {\bibfnamefont {H.}~\bibnamefont {Fariborzi}},\ }\bibfield  {title}
  {\bibinfo {title} {Near-zero stiffness accelerometer with buckling of tunable
  electrothermal microbeams},\ }\href@noop {} {\bibfield  {journal} {\bibinfo
  {journal} {Microsystems \& Nanoengineering}\ }\textbf {\bibinfo {volume}
  {10}},\ \bibinfo {pages} {43} (\bibinfo {year} {2024})}\BibitemShut {NoStop}%
\bibitem [{\citenamefont {Zhao}\ \emph {et~al.}(2017)\citenamefont {Zhao},
  \citenamefont {Li}, \citenamefont {Wang}, \citenamefont {Wang}, \citenamefont
  {Zhang},\ and\ \citenamefont {Gai}}]{zhao2017membrane}%
  \BibitemOpen
  \bibfield  {author} {\bibinfo {author} {\bibfnamefont {J.}~\bibnamefont
  {Zhao}}, \bibinfo {author} {\bibfnamefont {X.}~\bibnamefont {Li}}, \bibinfo
  {author} {\bibfnamefont {Y.}~\bibnamefont {Wang}}, \bibinfo {author}
  {\bibfnamefont {W.}~\bibnamefont {Wang}}, \bibinfo {author} {\bibfnamefont
  {B.}~\bibnamefont {Zhang}},\ and\ \bibinfo {author} {\bibfnamefont
  {X.}~\bibnamefont {Gai}},\ }\bibfield  {title} {\bibinfo {title} {Membrane
  acoustic metamaterial absorbers with magnetic negative stiffness},\
  }\href@noop {} {\bibfield  {journal} {\bibinfo  {journal} {The Journal of the
  Acoustical Society of America}\ }\textbf {\bibinfo {volume} {141}},\ \bibinfo
  {pages} {840} (\bibinfo {year} {2017})}\BibitemShut {NoStop}%
\bibitem [{\citenamefont {Zhang}\ \emph {et~al.}(2014)\citenamefont {Zhang},
  \citenamefont {Chan},\ and\ \citenamefont {Huang}}]{zhang2014thin}%
  \BibitemOpen
  \bibfield  {author} {\bibinfo {author} {\bibfnamefont {Y.}~\bibnamefont
  {Zhang}}, \bibinfo {author} {\bibfnamefont {Y.-J.}\ \bibnamefont {Chan}},\
  and\ \bibinfo {author} {\bibfnamefont {L.}~\bibnamefont {Huang}},\ }\bibfield
   {title} {\bibinfo {title} {Thin broadband noise absorption through acoustic
  reactance control by electro-mechanical coupling without sensor},\
  }\href@noop {} {\bibfield  {journal} {\bibinfo  {journal} {The Journal of the
  Acoustical Society of America}\ }\textbf {\bibinfo {volume} {135}},\ \bibinfo
  {pages} {2738} (\bibinfo {year} {2014})}\BibitemShut {NoStop}%
\bibitem [{\citenamefont {Yang}\ \emph {et~al.}(2014)\citenamefont {Yang},
  \citenamefont {Ma}, \citenamefont {Wu}, \citenamefont {Yang},\ and\
  \citenamefont {Sheng}}]{yang2014homogenization}%
  \BibitemOpen
  \bibfield  {author} {\bibinfo {author} {\bibfnamefont {M.}~\bibnamefont
  {Yang}}, \bibinfo {author} {\bibfnamefont {G.}~\bibnamefont {Ma}}, \bibinfo
  {author} {\bibfnamefont {Y.}~\bibnamefont {Wu}}, \bibinfo {author}
  {\bibfnamefont {Z.}~\bibnamefont {Yang}},\ and\ \bibinfo {author}
  {\bibfnamefont {P.}~\bibnamefont {Sheng}},\ }\bibfield  {title} {\bibinfo
  {title} {Homogenization scheme for acoustic metamaterials},\ }\href@noop {}
  {\bibfield  {journal} {\bibinfo  {journal} {Physical Review B}\ }\textbf
  {\bibinfo {volume} {89}},\ \bibinfo {pages} {064309} (\bibinfo {year}
  {2014})}\BibitemShut {NoStop}%
\bibitem [{\citenamefont {Yang}(2014)}]{yang2014metamaterial}%
  \BibitemOpen
  \bibfield  {author} {\bibinfo {author} {\bibfnamefont {M.}~\bibnamefont
  {Yang}},\ }\href@noop {} {\emph {\bibinfo {title} {Metamaterial
  homogenization and acoustic metasurface}}}\ (\bibinfo  {publisher} {Hong Kong
  University of Science and Technology (Hong Kong)},\ \bibinfo {year}
  {2014})\BibitemShut {NoStop}%
\bibitem [{\citenamefont {H{\"y}tch}\ and\ \citenamefont
  {Hawkes}(2022)}]{hytch2022quadrupoles}%
  \BibitemOpen
  \bibfield  {author} {\bibinfo {author} {\bibfnamefont {M.}~\bibnamefont
  {H{\"y}tch}}\ and\ \bibinfo {author} {\bibfnamefont {P.~W.}\ \bibnamefont
  {Hawkes}},\ }\href@noop {} {\emph {\bibinfo {title} {Quadrupoles in electron
  lens design}}}\ (\bibinfo  {publisher} {Academic Press},\ \bibinfo {year}
  {2022})\BibitemShut {NoStop}%
\bibitem [{\citenamefont {Rayleigh}(1929)}]{rayleigh1929theory-s}%
  \BibitemOpen
  \bibfield  {author} {\bibinfo {author} {\bibfnamefont {J.~W. S.~B.}\
  \bibnamefont {Rayleigh}},\ }\bibinfo {title} {Theory of sound}\ (\bibinfo
  {publisher} {MacMillan},\ \bibinfo {address} {New York},\ \bibinfo {year}
  {1929})\ p.\ \bibinfo {pages} {327}\BibitemShut {NoStop}%
\bibitem [{\citenamefont {Crandall}(1926)}]{crandall1926theory-s}%
  \BibitemOpen
  \bibfield  {author} {\bibinfo {author} {\bibfnamefont {I.}~\bibnamefont
  {Crandall}},\ }\bibinfo {title} {Theory of vibration system and sound}\
  (\bibinfo  {publisher} {Van Nostrand},\ \bibinfo {address} {New York},\
  \bibinfo {year} {1926})\ p.\ \bibinfo {pages} {229}\BibitemShut {NoStop}%
\bibitem [{\citenamefont {Ingard}(1953)}]{ingard1953theory-s}%
  \BibitemOpen
  \bibfield  {author} {\bibinfo {author} {\bibfnamefont {U.}~\bibnamefont
  {Ingard}},\ }\bibfield  {title} {\bibinfo {title} {On the theory and design
  of acoustic resonators},\ }\href@noop {} {\bibfield  {journal} {\bibinfo
  {journal} {The Journal of the acoustical society of America}\ }\textbf
  {\bibinfo {volume} {25}},\ \bibinfo {pages} {1037} (\bibinfo {year}
  {1953})}\BibitemShut {NoStop}%
\end{thebibliography}%

\end{document}